# Design Obligations for Software, with Examples from Data Abstraction and Adaptive Systems

Mary Shaw
School of Computer Science
Carnegie Mallon University
Pittsburgh, PA, USA
mary.shaw@cs.cmu.edu

*Abstract*—Producing a good software design involves not only writing a definition that satisfies the syntax of the chosen language or structural constraints of a design paradigm. It also involves upholding a variety of expectations about the behavior of the system—the semantic expectations. These expectations may apply not only at the code level, but also to more abstract system structures such as software architectures. Such high-level design paradigms provide a vocabulary of components or other constructs and ways to compose those constructs, but not all expressible designs are well-formed, and even well-formed designs may fail to satisfy the expectations of the paradigm.

Unfortunately, these expectations are often implicit or documented only informally, so they are challenging to discover, let alone uphold. They may for example, require correct use of complex structures, internal consistency, compliance with external standards, adherence with design principles, etc. Further, the reasons for design decisions that uphold these expectations are often not explicit in the code or other representation of the system. I introduce the idea of *design obligations*, which are constraints on allowable designs within a given design paradigm that help to assure appropriate use of the paradigm. To illustrate this idea, I discuss design obligations for two paradigms: data abstraction and a class of adaptive based on feedback control.

## I. Expectations Beyond the Syntax

Many aspects of software development and analysis have expectations—conventions, constraints, and rules—that shape the code or other representation of the software but are not explicit in the final product. The associated obligations are often not well understood, or at least not well documented. They are often related to the semantics rather than the syntactic correctness of the code. Yet the correctness or even adequacy of the software depends on responsible observance of these expectations. I argue here that these *design obligations* should be made explicit and supported more robustly.

For example, simple induction proofs have two proof obligations: a basis step in which you are obligated to show that the property of interest holds for an initial case, and an induction step in which you are obligated to show that if the property of interest holds for the k-th case it will necessarily hold for the k+1-th case. If you satisfy these two obligations, you have completed the proof [8].

The idea of design obligations is widespread in software engineering, usually by different names. These obligations arise from the programming language, application domain, or design paradigm itself. They provide software designers with responsibilities or issues to consider each time they develop a system. This section describes some common examples of these obligations. These are simply examples; this list is far from exhaustive.

Section II elaborates the idea of design obligations for high-level design, and Sections III and IV provide examples to make this concrete. Section III reformulates the well-known verification requirement for data abstraction as a set of design obligations, and Section IV derives obligations for a class of adaptive systems from a close inspection of elementary feedback control. Section V argues that software designing would benefit from making these obligations explicit.

### A. Programming Obligations

Programming practice has a long litany of *good practices*, many of which are obligations about usage that are not enforced by the programming language. These include guidance such as

- Don't use null pointers.
- Don't embed unsanitized user input in SQL queries.
- Be sure your units are consistent across modules (don't mix meters and feet).
- Be sure array indices are in range.

These are obligations about the values of variables or about consistency of shared assumptions. They go beyond things that can be enforced in the programming language or its type system.

### B. Representation Obligations

In database design, *integrity constraints* are protocols for data columns or tables. They ensure consistency and quality of data by restricting operations on the columns or tables. They may limit values, provide defaults, demand non-duplication, require unique primary keys, and so on [9].

### C. Usage Obligations

Libraries have expectations about usage. A UX designer found inconsistencies in the way library components were used. In identifying and documenting expectations for consistency, she learned that her UI library was a collection of components and styles, but she needed a design system with a set of *guidelines* about how the library elements were to be used [14].

Supported by the Alan J. Perlis professorship in Computer Science at Carnegie Mellon University.

### D. System-level Obligations

In unix, processes have independent memory address spaces, so one process cannot corrupt the memory of another process. In contrast, multiple threads can share an address space, so one thread can corrupt the memory of another thread. However, processes incur more overhead than threads. A programmer who chooses threads, for example to gain efficiency, accepts the *responsibility* for ensuring that the threads do not interfere with each other's memory. In other words, a unix programmer is obligated to clearly consider the question, "what is the blast radius of memory corruption?", and the answer will guide the process vs thread decision [10].

### E. Abstraction Obligations

Adding a layer of abstraction in a system creates an *abstraction obligation* to show that the abstraction is consistent with the underlying representation. That is, you must show that

- There is a well-defined mapping between the abstraction and the implementation.
- The effect of an operation in the abstract domain matches the effect actually achieved by the implementation.
- Everything that the abstraction can describe is properly handled in the implementation.

Section III discusses the special case of data abstractions.

### F. Requirements obligations

A system should meet all its requirements, and it should be possible to show how various elements of the implementation together meet these requirements. Similarly, all the elements in the implementation should arguably support some of the requirements. That is, there should be *traceability* between the requirements and the implementation.

### G. Architectural Obligations

Software architectures are described in terms of components and connectors of the software, but most architectural styles have further expectations. For example, the pattern template for pattern-oriented software architecture includes a section that describes the problem addressed by the pattern [3]. In addition to capturing the essence of the design problem, this section documents the *forces* operating on the problem. These forces identify aspects of the problem that should be considered in the solution. They might be, for example, performance requirements, constraints about allowable protocols, or desirable properties of solutions. They may be in competition with each other, thereby forcing consideration of design tradeoffs.

For example, forces for the pipe and filter architecture include the desirability of small processing steps, the expectation that non-adjacent filters do not share information, and the error-prone-ness of explicit storage of intermediate results by users (which motivates the use of pipes for communication).

## II. Design Obligations

*Design obligations* are design-level responsibilities that are analogous to proof obligations, good coding practices, integrity constraints, guidelines, and the like. For example, proving loop invariants in a programming language requires inductive proofs.

Similarly, some architectures rely on specific relations among components that impose design obligations to establish and preserve these critical relations. For example, databases rely on integrity constraints, and event systems require the designer to ensure that the collection of event handlers will, in aggregate, properly field all the events that are raised.

Hoare laid out the proof obligations for the representations of abstract data types [5]; this is the example of Section III. In feedback-based adaptive control, a set of obligations arise from the relations that must be assured among elements of the feedback process; this is the example of Section IV.

At all software levels, including code, architecture, design, and requirements, there are languages or language-like constructs that guide allowable design representations. These do not generally include all constraints that assure that the definitions will fully satisfy the expectations of the paradigm. The additional constraints, or expectations, are often tacit; they may sometimes be inferred from API documentation. Nevertheless, they create obligations on developers beyond the languages' rules on what can be defined[1].

I'm interested here in *design obligations*, the additional expectations that shape whether designs that are well-formed in a language or paradigm will actually be satisfactory. These are obligations that arise from the domain of the design paradigm itself; they are not specifications or rationales for particular designs developed in the paradigm.

Accordingly, the design obligations should refer only to the design paradigm. Explanations of the rationale for specific design decisions in a particular implementation and specifications of application-specific policies, for example with session contracts, is also important—but it's a different undertaking; the design obligations are one source of issues to address in those specifications.

## III. Example: Design Obligations for Data Abstractions

Whenever designers decide to approach a problem by introducing an abstraction, they incur an obligation to establish consistency between the abstraction and the system it abstracts from (as discussed in section I.E). Hoare describes one important approach, giving the design obligations for the abstraction layer associated with abstract data types [5]. This technique applies as well to other abstractions, for example transform spaces, simulations, model-driven design, and higher-level languages.

An abstract data type with a specification at the level of the abstraction is written in some specification and programming languages; this alone does not ensure that the implementation actually supports the abstraction correctly. Hoare lays out three design obligations for such abstract data types.

---

[1] I take a broad view of "programming language" here: a programming language is a symbolic notation with (a) a defined syntax; (b) language constructs including primitive elements, expressions for computing new elements, abstraction mechanisms for composing expressions, and perhaps a closure rule that compositions of expressions must yield legal elements; and (c) semantics that associate meaning in the application space with programs (after [1]). Even in high-level design, there are primitives, constructs, and compositions; the designer needs to understand the additional responsibilities associated with satisfying the expectations.

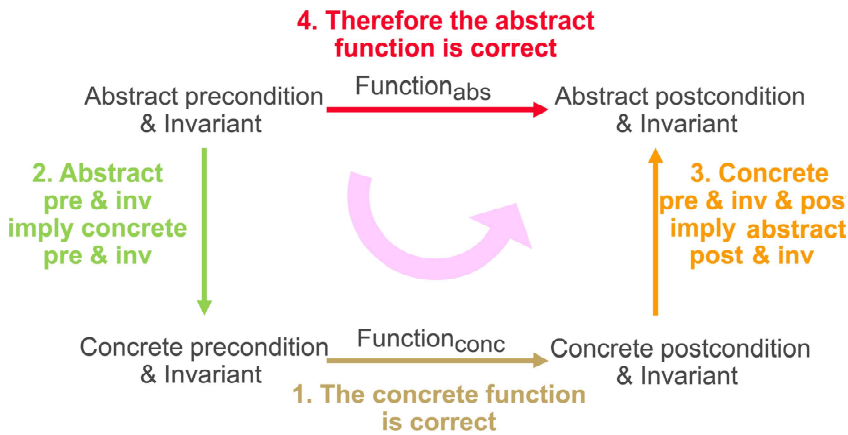

Fig. 2. Hoare's proof obligations for data abstraction

*Define an invariant for the representation.* This is an integrity constraint: a predicate that says whether the values in the implementation represent a legal value. For example, a stack implemented as a vector and an index into the vector would have an invariant that limits the index to values within the vector. During execution of a concrete function, the invariant may be violated, but it must be re-established before the end of the concrete function. The invariant is included as part of the pre-condition and post-condition in verification.

*Define an abstraction function.* This function maps between the representation and the abstraction: that is, it defines the relation of a (legal) value in the representation to an abstract value. In the case of the stack, it would give an abstract stack with only as many elements as the index indicates. This is represented by the vertical arrows in Figure 1.

*Prove three theorems about each operation*, as indicated by the numbered comments in Figure 1.

#1: Verify that the concrete function $Function_{cons}$ is correct.

#2: Verify that the abstract precondition implies the precondition of the concrete function, using the abstraction function.

#3: Verify that the concrete postcondition implies the abstract postcondition, using the abstraction function.

Those three theorems are sufficient to establish that the abstract function $Function_{abs}$ (#4) is correct, as suggested by the central arrow of Figure 1. Or, as the algebraists say "hurrah, the diagram commutes!"

## IV. EXAMPLE: DESIGN OBLIGATIONS FOR ADAPTIVE SYSEMS

Adaptive systems have a target criterion that specifies some aspect of its state during proper operation. They respond to changes in their process or operating environment by noticing the changes and altering the operating parameters of the process to maintain the target criterion. The relations among the environment, the state of the process, and the operating parameters of the process can be complex, perhaps even nonlinear. This means that the task of deciding how much to adjust the operating parameters in response to a change in the environment or the process can be quite complex.

The adaptation task resembles the feedback loops of control theory [2]. Figure 2 shows a simple control loop. A process $P$ is intended to operate at a reference level $y_r$; this is the target criterion. The state of the process $y$ is sensed periodically or continuously, possibly through a transducer, and the reported value of $y$ is compared to $y_r$ to determine the control error. A controller $C$ determines a control input $u$ to $P$; the algorithm of $C$ is responsible for selecting the right value of the correction $u$ to keep $y$ close enough to $y_r$.

This sort of feedback control is well understood; complete analysis and guarantees on performance can be assured. Unfortunately, the price of this assurance is that they are limited to simple relations.

In modern adaptive software the control relations are richer, so the complete analysis isn't feasible. The properties that affect whether the process is fully under control are not obvious, and they're not among the relations that can be described in UML [11][12]. As a result, the usual development processes do not call the designer's attention to these questions. The commonly-used MAPE architecture for adaptive systems also does not call for designers to be explicit about how they actually achieve control [7].

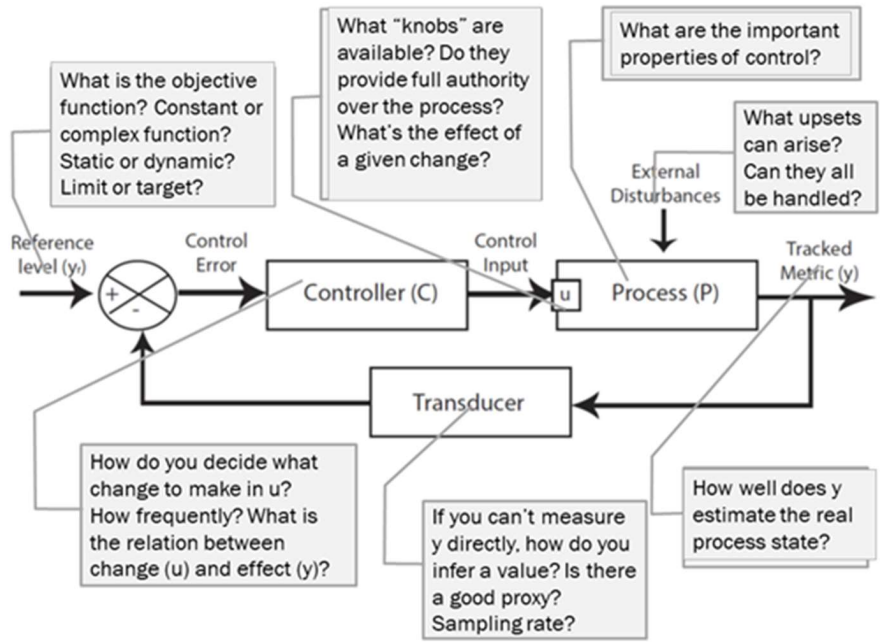

Fig. 1. Design obligations for adaptive control [13]

It is valuable to make design obligations explicit—but what *are* these obligations here? We can identify some of them by looking carefully at the structure of feedback control to find obligations that apply also to higher-level adaptive control. For example, the designer of an adaptive system that is supposed to maintain some parameter at a reference value should show that the system is controllable (the reference value is achievable), stable (control actions do move the system to the reference value), and robust (it's stable despite external disturbances).

A Dagstuhl on self-adaptive systems [4] explored in depth the application of control theory to self-adaptive systems. In particular, we identified the limitations of classical control theory for large-scale adaptive systems, showed how aspects of classical control could nonetheless guide design, and worked through several case studies [6]. These ideas are perhaps more accessible in a related video [13], which examines the components of a feedback loop and identifies design obligations—properties that the designer must address—as shown in the boxed annotations in Figure 2. They call out the need for the designer to consider

- what are the important properties to control
- how the reference level is determined
- how well the tracked metric estimates the real process state, and what proxy might suffice if direct measurement isn't possible
- what operations on the process are available, and are they adequate for full control
- how the corrections are determined, and what effect a correction has on the process
- whether the available operations on the process are sufficient to maintain control
- what external disturbances are possible, and whether it is possible to compensate for them with process inputs

These are essential questions that the usual software development methods and tools, do not address.

## V. Conclusion

It isn't sufficient to write code or lay out a design in whatever notation is available. When designing within a framework, paradigm, or other norm, the designer often has additional obligations to ensure certain properties of the design. Unfortunately, these are often informal or implicit.

This position paper calls out the need for more systematic, even more formal, recognition and documentation of these obligations. Doing so would benefit both the learning and the checking of these obligations.

## Acknowledgment

The description of design obligations for adaptive systems draws heavily on discussions at Dagstuhl Seminars on adaptive systems, especially with Marin Litoiu, Gabriel Tamura, Norha Villegas, Hausi Mueller, and Mauro Pezzè, I also appreciate advice and feedback from Josh Bloch, André van der Hoek, Marian Petre, Mahadev Satyanarayanan, Dave Thomas, and the CMU software engineering group, especially George Fairbanks, Claire Le Goues, Heather Miller, Ipek Ozkaya, and Roy Weil.